\documentclass[a4paper,10pt]{paper}
\usepackage[utf8x]{inputenc}

\usepackage{amsmath}
\usepackage{amsfonts}
\usepackage[retainorgcmds]{IEEEtrantools}
\usepackage[noblocks]{authblk}
\newtheorem{theorem}{Theorem}

\newtheorem{corollary}[theorem]{Corollary}

\newtheorem{lemma}[theorem]{Lemma}

\newtheorem{remark}[theorem]{Remark}

\newcommand{\hr}{{\mathcal H}}

\newcommand{\cs}{{\mathcal S}}

\newcommand{\kr}{{\mathcal K}}

\newcommand{\cc}{{\mathbb C}}
\newcommand{\rr}{{\mathbb R}}
\newcommand{\nn}{{\mathbb N}}
\newcommand{\idn}{\mathbf{1}}

\newcommand{\eps}{{\varepsilon}}        

\newcommand{\cl}{\mathcal L}
\newcommand{\tr}{\textrm{tr}}

\title{Quantum Stein's lemma revisited, inequalities for quantum entropies, and a concavity theorem of Lieb}
\author[1]{Igor Bjelakovi\'c}
\author[2]{Rainer Siegmund-Schultze}
\small \affil[1]{Technische Universit\"at M\"unchen, Theoretische Informationstechnik,\authorcr
Arcisstra\ss e 21,
80333 M\"unchen, Germany\authorcr igor.bjelakovic@tum.de}
\affil[2]{Technische Universit\"at Berlin, Institut f\"ur Mathematik,\authorcr Stra\ss e des 17. Juni 136,
10623 Berlin, Germany\authorcr siegmund@math.tu-berlin.de
}
\begin{document}
\maketitle

\begin{abstract}
We derive the monotonicity of the quantum relative entropy by an elementary operational argument based on
Stein's lemma in quantum hypothesis testing. For the latter we present an elementary and short proof that requires the law of large numbers only.  
Joint convexity of the quantum relative entropy is proven too, resulting in a self-contained 
elementary version of Tropp's approach to Lieb's concavity theorem, 
according to which the map 
$a\mapsto \tr(\exp(h+\log a))$ is concave 
on positive operators for self-adjoint $h$. 
\end{abstract}
\tableofcontents
\section{Introduction}
Inequalities for quantum mechanical entropies and related concave trace functions play a fundamental role
in quantum information theory. The ground-breaking results on these inequalities by Lieb \cite{lieb} and by Lieb and Ruskai \cite{lieb-ruskai}
together with the extension of the fundamental operational ideas and concepts from Shannon's information theory \cite{shannon} to the quantum realm 
have made the rapid development of quantum information theory possible. For example, many optimality proofs in quantum
information theory rely on one of the fundamental inequalities established in \cite{lieb} and \cite{lieb-ruskai}. The proofs
presented in \cite{lieb} and \cite{lieb-ruskai} are masterpieces of matrix analysis.\\
In this paper we approach some of the major inequalities for quantum entropies from the point of view of Shannon's theory. 
It turns out that the monotonicity property of the quantum
relative entropy \cite{lieb-ruskai, lindblad-cptp-mono, uhlmann-wyd-concavity} is an elementary and intuitive consequence of 
the quantum version of Stein's lemma \cite{hiai-petz, ogawa}, which gives an operational interpretation
to the quantum relative entropy as a distinguishability measure on the set of quantum states. 
Stein's lemma, in turn, can be established by an elementary argument on less than two pages: The new proof that we present below requires only 
the law of large numbers and a simple estimate on ``overlaps'' of certain projections with respect to some given quantum state. It is this simple proof
of Stein's lemma that we consider as the main contribution of the paper. The observation that the monotonicity of the relative entropy can be derived
from Stein's lemma has been made in the technical report \cite{mono-1} by the authors almost a decade ago. There we have given an elementary but,
 unfortunately, somewhat non-transparent proof of Stein's lemma.\\
Once the monotonicity is established, we have access to other fundamental properties of quantum entropies as described in Ruskai's review 
\cite[Sec.~V]{ruskai}. We restrict our attention to the joint convexity of the quantum relative entropy \cite{lindblad, uhlmann-wyd-concavity}
for which we include a short proof. Indeed, the joint convexity and monotonicity of the relative entropy are equivalent \cite[Sec.~V]{ruskai} and 
therefore we could easily obtain it directly from Stein's lemma. We prefer, however, to follow the nice derivation from the monotonicity mentioned in
\cite[Sec.~V]{ruskai}.\\
Our motivation for including the joint convexity of relative entropy is the recent work \cite{tropp} by Tropp where he derives one of Lieb's 
concavity theorems \cite{lieb}, 
according to which the map $a\mapsto\tr(\exp(h+\log a))$ is concave on the positive cone for fixed self-adjoint $h$, from the joint convexity 
of the relative entropy. We present a short and slightly streamlined version of Tropp's argument in Section \ref{sec:tropp}. Together with our
operational derivation of the monotonicity of the quantum relative entropy we obtain a self-contained access to some of the fundamental inequalities
for quantum entropies and trace functions from the information-theoretic perspective.\\ 
Besides the fact that Lieb's concavity theorem played a crucial role in \cite{lieb-ruskai} it became of great importance for establishing
sharp tail concentration bounds for sums of random matrices \cite{tropp-tail}. The latter development, in turn, 
has its origin in the famous Ahlswede-Winter bound \cite{ahlswede-winter}
that arose in the context of the theory of identification via quantum channels.\\
We resisted the temptation of producing an extremely short paper, which could be done given the elementary nature of the arguments that are used. 
Instead, our leitmotif was to give a self-contained presentation of the results at a slow pace so that anybody knowing the law of large numbers and
being familiar with basic linear algebra can easily follow our arguments. Only exception being Remark \ref{remark-general-mono} 
in Section \ref{sec:mono} where
we assume the familiarity with the definition and simple properties of completely positive maps which can be easily 
picked up in Bhatia's beautiful book \cite[Ch.~3]{bhatia}. We should note, however, that no result in the paper depends on the inequality
presented in Remark \ref{remark-general-mono}. It is included for completeness only and can safely be skipped without any 
consequence for the subsequent parts of the paper.  
\section{An elementary proof of quantum Stein's lemma}\label{sec:stein}

For $\rho,\sigma\in \cs(\hr)$\footnote{$\cs(\hr):=\{\rho\in\cl(\hr): \rho\ge 0, \tr(\rho)=1  \}$ denotes the set of density 
operators or states on $\hr$ whereas $\cl(\hr)$ stands for the set of linear maps from $\hr$ to itself.},
 $\eps\in (0,1)$, and $n\in\nn$ we define
\begin{equation}\label{eq:sep-number}
 \beta_{\eps,n}(\rho,\sigma):=\min\left\{ \tr (\sigma^{\otimes n}a): a\in [0,\idn_{\hr^{\otimes n}}], 
\tr (\rho^{\otimes n} a)\ge 1-\eps  \right\},
\end{equation} 
where $ [0,\idn_{\hr^{\otimes n}}]$ denotes the set of (self-adjoint) operators $a$ on $\hr^{\otimes n}$ with $0\le a\le\idn_{\hr^{\otimes n}}$.\\
The quantities $\beta_{\eps,n}(\rho,\sigma)$ obtain their natural interpretation in terms of statistical
hypothesis testing. We suppose that the system under consideration is prepared either according to the state $\rho$
or to the state $\sigma$ and we can perform measurements/observations on $n$ independently prepared systems whose state
is then $\rho^{\otimes n}$ or $\sigma^{\otimes n}$. 
According to Quantum Mechanics, a binary observable of the $n$-partite system is a map
$E:\{ 0,1 \}\to [0,\idn_{\hr^{\otimes n}}]$ such that $E(0)+E(1)=\idn_{\hr^{\otimes n}}$. Given a state $\tau\in\cs(\hr^{\otimes n})$
Quantum Mechanics assigns the probabilities 
\begin{equation}\label{probab-assig}
 \tr(\tau E(i))\in [0,1]\qquad (i\in \{ 0,1 \}) 
\end{equation}   
for obtaining the outcome $i$ when measuring the observable $E$ and when the system is prepared in the state $\tau$. 
Notice that this is consistent with the requirement  
$E(0)+E(1)=\idn_{\hr^{\otimes n}}$ leading to $\tr(\tau E(0))+\tr(\tau E(1))=1$ for all $\tau\in\cs(\hr^{\otimes n})$.\\
Given states $\rho,\sigma\in \cs(\hr)$ the probabilities for obtaining the outcome $0$ for the observable $E$ 
are given by $\tr(\rho^{\otimes n} a)$ and $\tr(\sigma^{\otimes n} a)$ with $a:=E(0)$ given that the $n$-partite system is prepared
either in state $\rho^{\otimes n}$ or in state $\sigma^{\otimes n}$.\\
Suppose that we use the observable $E$ as a decision rule, meaning that when obtaining the outcome
$0$ we decide that the state was $\rho$ and else we decide in favor of $\sigma$. Then $\tr(\sigma^{\otimes n} a)$ (again abbreviating $a=E(0)$)
represents the error probability of our decision rule and the numbers $\beta_{\eps,n}(\rho,\sigma)$ are
the minimum error probabilities for the decision rules deciding in favor  of $\rho$ with high probability (assuming that $\eps$ is 
close to $0$).\\
The quantum relative entropy of $\rho,\sigma\in\cs(\hr)$ is given by\footnote{All logarithms are to the base $e$.}
\begin{equation}\label{def-rel-entropy}
 D(\rho||\sigma):=
\begin{cases}
 \tr(\rho\log\rho-\rho\log\sigma)& \text{if } \ker\sigma\subseteq \ker\rho\\
 +\infty & \text{else}.
\end{cases}
\end{equation}  
\begin{theorem}[Quantum Stein's lemma \cite{hiai-petz}, \cite{ogawa}]\label{theorem-stein's-lemma}
Let $\rho,\sigma \in \cs(\hr)$ be states with $\ker \sigma\subseteq \ker \rho$. Then for any $\eps\in (0,1)$ we have
\begin{equation}\label{stein's-lemma}
 \lim_{n\to\infty}\frac{1}{n}\log \beta_{\eps,n}(\rho,\sigma)=-D(\rho||\sigma).
\end{equation} 
\end{theorem}
\begin{remark}  Stein's Lemma shows that, roughly, $\beta_{\eps,n}(\rho,\sigma)\approx e^{-n D(\rho||\sigma)}$ giving an operational interpretation
to the quantum relative entropy as the largest rate at which the error probability decays to $0$ exponentially fast.
\end{remark}
The proof of Theorem \ref{theorem-stein's-lemma} that is presented below relies on two simple 
lemmas that are proven first.\\
Let $\tau_1,\tau_2\in \cs(\hr)$ and assume that $\ker \tau_2 \subseteq \ker\tau_1$ holds. We set
\begin{equation}\label{def-M}
 M(\tau_1||\tau_2):=-\tr (\tau_1\log \tau_2).
\end{equation} 
\begin{remark}
 Notice that
\begin{equation}\label{m-rel-entropy}
 M(\tau_1||\tau_2)-S(\tau_1)=D(\tau_1||\tau_2),
\end{equation} 
and
\begin{equation}\label{m-entropy}
 M(\tau_1||\tau_1)=S(\tau_1)
\end{equation} 
hold, where $S(\tau_1):=-\tr(\tau_1\log \tau_1)$ is the von Neumann entropy of $\tau_1$.
\end{remark}

Let $\mu_1,\ldots,\mu_d$ be the eigenvalues of $\tau_2$ counted with their multiplicities and $e_1,\ldots,e_d$ a complete
orthonormal set of corresponding eigenvectors. For $[ d ]:=\{1,\ldots, d   \}$ and $x^n:=(x_1,\ldots,x_n)\in [  d]^n$ we set
\begin{equation}
 \mu_{x^n}:=\prod_{i=1}^{n}\mu_{x_i}\qquad e_{x^n}:=\bigotimes_{i=1}^{n}e_{x_i}.
\end{equation} 
Additionally, for $\delta>0$ we set
\begin{IEEEeqnarray}{rCl}\label{def-typical-set}
T_{\delta,n}(\tau_1,\tau_2)&:=&\left\{ x^n\in [d]^n: M(\tau_1||\tau_2)-\delta< -\frac{1}{n}\log \mu_{x^n}<  M(\tau_1||\tau_2)+\delta
 \right\}\\
&=& \left\{x^n\in [d]^n: e^{-n(M(\tau_1||\tau_2)+\delta)}<\tr (\tau_2^{\otimes n}|e_{x^n}\rangle\langle e_{x^n}|)< e^{-n(M(\tau_1||\tau_2)-\delta)}
 \nonumber
 \right\}.
\end{IEEEeqnarray}
Finally we introduce the following projection 
\begin{equation}\label{def-typical-projector}
 p_{\delta,n}(\tau_1,\tau_2):=\sum_{x^n\in T_{\delta,n}(\tau_1,\tau_2)} |e_{x^n}\rangle\langle e_{x^n}|.
\end{equation} 
\begin{lemma}\label{key-lemma}
For all $\tau_1,\tau_2\in\cs(\hr)$ with $\ker\tau_2\subseteq \ker\tau_1$ and all $\delta>0$ we have:
\begin{enumerate}
 \item $p_{\delta,n}(\tau_1,\tau_2)\tau_2^{\otimes n}= \tau_2^{\otimes n} p_{\delta,n}(\tau_1,\tau_2)$ for all $n\in\nn$.
 \item $ p_{\delta,n}(\tau_1,\tau_2)\tau_2^{\otimes n} p_{\delta,n}(\tau_1, \tau_2)\le  e^{-n(M(\tau_1||\tau_2)-\delta)}
p_{\delta,n}(\tau_1,\tau_2) $ for all $n\in\nn$
\item $ p_{\delta,n}(\tau_1,\tau_2)\tau_2^{\otimes n} p_{\delta,n}(\tau_1, \tau_2)\ge  e^{-n(M(\tau_1||\tau_2)+\delta)}
p_{\delta,n}(\tau_1,\tau_2) $ for all $n\in\nn$
\item $\lim_{n\to\infty}\tr (\tau_1^{\otimes n} p_{\delta,n}(\tau_1,\tau_2))=1$.
\end{enumerate}
\end{lemma}
{\sf{Proof:}} The first three claims in the lemma are obvious from the definition of $p_{\delta,n}(\tau_1,\tau_2)$.
The last assertion follows from the law of large numbers: First of all, we can w.l.o.g. assume that $\tau_2$ is invertible
due to our assumption that $\ker\tau_2\subseteq\ker\tau_1$.
Let $X_1,\ldots, X_n$ be independent, identically distributed (i.i.d.)
random variables taking values in $[d]$ with distribution
\begin{equation}\label{pre-random-variables}
 \Pr (X_1=x_1,\ldots, X_n=x_n)=\tr (\tau_1^{\otimes n}|e_{x^n}\rangle\langle e_{x^n}|)=\prod_{i=1}^n\tr(\tau_1 |e_{x_i}\rangle\langle e_{x_i}|).
\end{equation} 
For $i=1,\ldots,n$ we introduce
\begin{equation}\label{random-variables}
 U_i:= -\log \mu_{X_i}
\end{equation} 
i.e. $U_i=f\circ X_i$ with the function $f:[d]\to\rr$, $f(x)=-\log \mu_x$.\\
$U^n:=(U_1,\ldots,U_n)$ is an i.i.d. collection of random variables and 
\begin{equation}\label{expectation}
 \mathbb{E}(U_i)=\sum_{x\in [d]}\tr(\tau_1 |e_x\rangle\langle e_x|)(-\log \mu_x)=M(\tau_1||\tau_2)
\end{equation} 
for all $i=1,\ldots, n$. Moreover, it is clear that
\begin{IEEEeqnarray}{rCl}\label{typical-relation}
\tr(\tau_1^{\otimes n} p_{\delta,n}(\tau_1,\tau_2))&=&\Pr(U^n\in T_{\delta,n}(\tau_1,\tau_2))\nonumber\\
&=&\Pr\left(\left|\sum_{i=1}^n U_i-n M(\tau_1||\tau_2)\right|<n\delta  \right)
\end{IEEEeqnarray}
Eqn. (\ref{expectation}), (\ref{typical-relation}), and the law of large numbers imply that
\begin{equation}
 \lim_{n\to\infty}\tr (\tau_1^{\otimes n} p_{\delta,n}(\tau_1,\tau_2))=1,
\end{equation} 
 as desired. \begin{flushright}$\Box$\end{flushright}
\begin{remark}\label{sm-projection}
In the proof of Theorem \ref{theorem-stein's-lemma} we will have to apply Lemma \ref{key-lemma} for the pairs $(\tau_1,\tau_2)=(\rho,\sigma)$ 
and $(\tau_1,\tau_2)=(\rho,\rho)$ simultaneously.
For the latter case we introduce a separate projection:
\begin{equation}\label{typ-sm-projection}
 p_{\delta,n}(\rho):=p_{\delta,n}(\rho,\rho).
\end{equation} 
\end{remark}
The next lemma is a fusion and a slight generalization of Lemma 6 in \cite{hayashi-nagaoka} and Lemma 8 in \cite{bdksss}. The proof is a standard
application of the Cauchy-Schwarz inequality for the Hilbert-Schmidt inner product on the space of linear operators over a finite-dimensional Hilbert 
space and is relegated to the Appendix \ref{proof-overlap-lemma}.
\begin{lemma}\label{overlap-lemma}
 Let $\kr$ be a Hilbert space over $\cc$ with $\dim \kr <\infty $.
Let $p,q\in \cl(\kr)$ with $0\le p,q\le\idn$ and $\tau \in \cs(\kr)$. Then
\begin{enumerate}
 \item $\tr (\tau pqp)\ge \tr(\tau q)-2\sqrt{\tr (\tau (\idn_{\kr}-p))}$.
 \item If $u\in\cl(\kr)$ is any projection commuting with $\tau$ (i.e. $u\tau=\tau u$) and satisfying
  $\tau u\le c u$ for some $c\in\rr_{+}$ then
 \[ \tr(pqp)\ge \frac{1}{c}\left( \tr(\tau q)-2\sqrt{\tr (\tau (\idn_{\kr}-p))}-\tr(\tau(\idn_{\kr}-u))  \right). \] 
\end{enumerate}
\end{lemma}
{\sf{Proof of Theorem \ref{theorem-stein's-lemma}:}} In a first step we show that for all $\eps\in (0,1)$
\[ \limsup_{n\to\infty}\frac{1}{n}\log \beta_{\eps,n}(\rho,\sigma)\le-D(\rho||\sigma). \]
Let $\delta>0$ be given. Then on account of Lemma \ref{key-lemma}.4 applied simultaneously to $(\tau_1,\tau_2)=(\rho,\sigma)$
and $(\tau_1,\tau_2)=(\rho,\rho)$, Remark \ref{sm-projection}, and Lemma \ref{overlap-lemma}.1 for any $\eps\in (0,1)$
 there is $n_0(\eps)\in \nn$ such that for all $n\ge n_0(\eps)$ we have
\begin{IEEEeqnarray}{rCl}\label{qsl-1}
 \tr(\rho^{\otimes n}p_{\delta,n}(\rho,\sigma)p_{\delta,n}(\rho)p_{\delta,n}(\rho,\sigma))
                                        &\ge& \tr(\rho^{\otimes n}p_{\delta,n}(\rho))\nonumber\\
                                              && -2\sqrt{\tr(\rho^{\otimes n}(\idn_{\hr^{\otimes n}}-
                                               p_{\delta,n}(\rho,\sigma)))}\nonumber\\
                                        &>&1-\eps.
\end{IEEEeqnarray} 
On the other hand, Lemma \ref{key-lemma}.1 and \ref{key-lemma}.2 imply ($(\tau_1,\tau_2)=(\rho,\sigma)$)
\begin{equation}\label{qsl-2}
 \sigma^{\otimes n}p_{\delta,n}(\rho,\sigma)\le e^{-n(M(\rho||\sigma)-\delta)}p_{\delta,n}(\rho,\sigma)
\end{equation} 
and consequently
\begin{equation}\label{qsl-3}
 p_{\delta,n}(\rho)\sigma^{\otimes n}p_{\delta,n}(\rho,\sigma)p_{\delta,n}(\rho)\le e^{-n(M(\rho||\sigma)-\delta)}
p_{\delta,n}(\rho)p_{\delta,n}(\rho,\sigma)p_{\delta,n}(\rho).
\end{equation} 
Taking trace in (\ref{qsl-3}) and observing that $\sigma^{\otimes n}$ and $p_{\delta,n}(\rho,\sigma)$ commute 
(cf. Lemma \ref{key-lemma}.1 with $(\tau_1,\tau_2)=(\rho,\sigma)$) we obtain
\begin{IEEEeqnarray}{rCl}\label{qsl-4}
 \tr(\sigma^{\otimes n}p_{\delta,n}(\rho,\sigma)p_{\delta,n}(\rho)p_{\delta,n}(\rho,\sigma) )&\le& 
e^{-n(M(\rho||\sigma)-\delta)} \tr (p_{\delta,n}(\rho)p_{\delta,n}(\rho,\sigma)p_{\delta,n}(\rho))\nonumber \\
&\le& e^{-n(M(\rho||\sigma)-\delta)} \tr(p_{\delta,n}(\rho))\nonumber\\
&\le& e^{-n(M(\rho||\sigma)-\delta)} e^{n(S(\rho)+\delta)}\tr (\rho^{\otimes n}p_{\delta,n}(\rho))\nonumber\\
&\le& e^{-n(D(\rho||\sigma)-2\delta )},
\end{IEEEeqnarray} 
where in the second line we have used
\begin{equation}\label{qsl-5}
 \tr (p_{\delta,n}(\rho)p_{\delta,n}(\rho,\sigma)p_{\delta,n}(\rho))\le\tr ( p_{\delta,n}(\rho))
\end{equation} 
which follows from $p_{\delta,n}(\rho,\sigma)\le \idn_{\hr^{\otimes n}}$ and $(p_{\delta,n}(\rho))^2=p_{\delta,n}(\rho)$. 
In the third line we have used Lemma \ref{key-lemma}.3 with 
$(\tau_1,\tau_2)=(\rho,\rho)$ together with $M(\rho||\rho)=S(\rho)$. The final line holds because $M(\rho||\sigma)-S(\rho)=D(\rho||\sigma)$
and $\tr(\rho^{\otimes n}p_{\delta,n}(\rho))\le 1$.\\
Defining 
\begin{equation}\label{qsl-6}
 a_n:= p_{\delta,n}(\rho, \sigma)p_{\delta,n}(\rho)p_{\delta,n}(\rho,\sigma)\in [0,\idn_{\hr^{\otimes n}}]
\end{equation} 
we obtain from (\ref{qsl-1}) and (\ref{qsl-4}) that 
\begin{equation}\label{qsl-7}
 \limsup_{n\to\infty}\frac{1}{n}\log \beta_{\eps,n}(\rho,\sigma)\le \limsup_{n\to\infty}\frac{1}{n}\log 
\tr(\sigma^{\otimes n}a_n)\le -D(\rho||\sigma)+2\delta.
\end{equation} 
Since $\delta>0$ is arbitrary we can conclude that
\begin{equation}\label{qsl-8}
 \limsup_{n\to\infty}\frac{1}{n}\log \beta_{\eps,n}(\rho,\sigma)\le-D(\rho||\sigma).
\end{equation}
We turn now to the proof of
\[ \liminf_{n\to\infty} \frac{1}{n}\log \beta_{\eps,n}(\rho,\sigma)\ge-D(\rho||\sigma).\] 
Let $\eps\in (0,1)$ be arbitrary and let 
\begin{equation}\label{qsl-9}
 q_n:=\arg\min \beta_{\eps,n}(\rho,\sigma).
\end{equation} 
From Lemma \ref{key-lemma}.1 and Lemma \ref{key-lemma}.3 with $(\tau_1,\tau_2)=(\rho,\sigma)$ we can infer that
\begin{IEEEeqnarray}{rCl}\label{qsl-10}
 \sigma^{\otimes n}&\ge& \sigma^{\otimes n}p_{\delta,n}(\rho,\sigma)\nonumber\\
                   &\ge& e^{-n(M(\rho||\sigma)+\delta)} p_{\delta,n}(\rho,\sigma)
\end{IEEEeqnarray} 
implying
\begin{equation}\label{qsl-11}
 q_n ^{1/2}\sigma^{\otimes n}q_n^{1/2}\ge e^{-n(M(\rho||\sigma)+\delta)}q_n^{1/2} p_{\delta,n}(\rho,\sigma) q_n^{1/2},
\end{equation} 
and taking the trace
\begin{IEEEeqnarray}{rCl}\label{qsl-12}
 \beta_{\eps,n}(\rho,\sigma)&=& \tr(\sigma^{\otimes n}q_n)\quad (\textrm{cf. (\ref{qsl-9})})\nonumber\\
                             &\ge& e^{-n(M(\rho||\sigma)+\delta)} \tr(q_n p_{\delta,n}(\rho,\sigma))\nonumber\\
                             &=& e^{-n(M(\rho||\sigma)+\delta)}\tr (p_{\delta,n}(\rho,\sigma)q_np_{\delta,n}(\rho,\sigma)). 
\end{IEEEeqnarray} 
Here we have used the cyclicity of the trace in the first and second line and, additionally, in the last line that 
 $(p_{\delta,n}(\rho,\sigma))^2=p_{\delta,n}(\rho,\sigma)$ holds.\\
We will lower-bound the last term in (\ref{qsl-12}).
Recall that Lemma \ref{key-lemma}.1 and Lemma \ref{key-lemma}.2 in the case $(\tau_1,\tau_2)=(\rho,\rho)$ guarantee that
\begin{equation}\label{qsl-13}
 p_{\delta,n}(\rho)\rho^{\otimes n}=\rho^{\otimes n}p_{\delta,n}(\rho)\textrm{ and } \rho^{\otimes n}p_{\delta,n}(\rho)
\le e^{-n(S(\rho)-\delta)}p_{\delta,n}.
\end{equation} 
 Then Lemma \ref{overlap-lemma}.2 and eq. (\ref{qsl-9}) show that there is $n_1(\eps)\in \nn$ such that
\begin{IEEEeqnarray}{rCl}\label{qsl-14}
 \tr(p_{\delta,n}(\rho,\sigma)q_np_{\delta,n}(\rho,\sigma) )&\ge& e^{n(S(\rho)-\delta)}(\tr(\rho^{\otimes n}q_n)
                                                              - 2\sqrt{\tr(\rho^{\otimes n}(\idn_{\hr^{\otimes n}}-p_{\delta,n}(\rho,\sigma)) )} \nonumber\\
                                                               & & -\tr(\rho^{\otimes n}(\idn_{\hr^{\otimes n}}-p_{\delta,n}(\rho))))\nonumber\\
                                                             &\ge& e^{n(S(\rho)-\delta)}\cdot \frac{1-\eps}{2}
\end{IEEEeqnarray}
for all $ n\ge n_1(\eps)$ by Lemma \ref{key-lemma}.4 applied to $p_{\delta,n}(\rho)$ and $p_{\delta,n}(\rho,\sigma)$.\\
The inequalities (\ref{qsl-12}),(\ref{qsl-14}), and $M(\rho||\sigma)-S(\rho)=D(\rho||\sigma)$ show that
\begin{equation}\label{qsl-15}
 \liminf_{n\to\infty} \frac{1}{n}\log \beta_{\eps,n}(\rho,\sigma)\ge-D(\rho||\sigma)-2\delta
\end{equation} 
holds for all $\delta>0$ and we are done.\begin{flushright}$\Box$\end{flushright}
\section{Monotonicity and joint convexity of quantum relative entropy}\label{sec:mono}
In this section we will show how Stein's lemma, Theorem \ref{theorem-stein's-lemma}, can be used to show the monotonicity
of the relative entropy under the partial trace as well as the joint convexity.\\
Let $\hr_1$, $\hr_2$ two finite-dimensional Hilbert spaces over $\cc$. The partial trace is given by the map
 $\tr_2:\cl(\hr_1)\otimes\cl(\hr_2)\to\cl(\hr_1)$, $\tr_2:=\textrm{id}_{\cl(\hr_1)}\otimes \tr_{\cl(\hr_2)}$, where
$\tr_{\cl(\hr_2)}$ denotes the trace on $\cl(\hr_2)$. Notice that for any state $\tau\in \cs(\hr_1\otimes\hr_2)$ and any
$a\in \cl(\hr_1)$
\begin{equation}\label{partial-trace-rel}
\tr_{\cl(\hr_1)}(\tr_2(\tau)a)=\tr_{\cl(\hr_1)\otimes\cl(\hr_2)}(\tau (a\otimes \idn_{\hr_2}))
 \end{equation}
holds. Introducing the map $E:\cl(\hr_1)\to \cl(\hr_1)\otimes \cl(\hr_2)$, $E(a):=a\otimes \idn_{\hr_2}$ this can be written
compactly as
\begin{equation}\label{partial-trace-1}
 \tr_{\cl(\hr_1)}(\tr_2(\tau)a)=\tr_{\cl(\hr_1)\otimes\cl(\hr_2)}(\tau E(a)).
\end{equation} 
Notice that the map $E$ has the following properties which are readily checked: For all $a,b\in \cl(\hr_1)$ we have
$ E(\idn_{\hr_1})=\idn_{\hr_1\otimes\hr_2},E(a^{\ast})=E(a)^{\ast}$,  and $E(ab)=E(a)E(b)$ and this already implies that $E$ is positive, 
i.e. preserves the positive
semi-definiteness of operators. However, the latter property is also obvious from the definition of $E$.\\ 
For $n\in\nn$ and all $\tau\in\cs((\hr_1\otimes \hr_2)^{\otimes n})$, $a\in\cl(\hr_1)^{\otimes n}$ we have 
\begin{equation}\label{partial-trace-2}
 \tr_{\cl(\hr_1)^{\otimes n}}(\tr_2^{\otimes n}(\tau)a)=\tr_{(\cl(\hr_1)\otimes \cl(\hr_2))^{\otimes n}}(\tau E^{\otimes n}(a)).
\end{equation} 
$E^{\otimes n}$ inherits the following properties: For all $a,b\in\cl(\hr_1)^{\otimes n}$
\begin{equation}\label{partial-trace-3}
 E^{\otimes n}(\idn_{\hr_1^{\otimes n}})= \idn_{(\hr_1\otimes\hr_2)^{\otimes n}}, \quad E^{\otimes n}(a^{\ast})=(E^{\otimes n}(a))^{\ast}, 
\end{equation}
and 
\begin{equation}\label{partial-trace-3-a} 
 E^{\otimes n}(ab)=E^{\otimes n}(a)E^{\otimes n}(b),
\end{equation} 
implying that $E^{\otimes n}$ is positive too\footnote{The properties in (\ref{partial-trace-3}) and (\ref{partial-trace-3-a}) are obvious on the set $G(n,\hr_1)
:=\{ \otimes_{i=1}^{n}a_i: a_i\in \cl(\hr_1), i=1,\ldots, n  \}$ and extend by linearity to the whole $\cl(\hr_1)^{\otimes n}$ since $G(n,\hr_1)$ is a 
generating set for $\cl(\hr_1)^{\otimes n}$. The positivity follows from $E^{\otimes n}(b^{\ast}b)=(E^{\otimes n}(b))^{\ast}E^{\otimes n}(b)$ 
and the fact that every positive semi-definite $a\in\cl(\hr_1)^{\otimes n}$ can be written as $a=b^{\ast}b$ for suitable $b\in\cl(\hr_1)^{\otimes n}$}.
From this we see that 
\begin{equation}\label{partial-trace-4}
P_n:=E^{\otimes n}([0,\idn_{\hr_1^{\otimes n}}])\subset [0,\idn_{(\hr_1\otimes \hr_2)^{\otimes n}}], 
\end{equation} 
where $[0,\idn_{\hr_1^{\otimes n}}]$ is the set of all (self-adjoint) operators $a\in \cl(\hr_1)^{\otimes n}$ with $0\le a\le \idn_{\hr_1^{\otimes n}}$
and $[0,\idn_{(\hr_1\otimes \hr_2)^{\otimes n}}]$ is defined correspondingly.
\begin{theorem}[Monotonicity under partial trace \cite{lieb-ruskai}]\label{mono-part-trace}
 Let $\rho,\sigma\in \cs(\hr_1\otimes\hr_2)$ be states. Then
\begin{equation}\label{eq:mono-part-trace}
 D(\rho||\sigma)\ge D(\tr_2 (\rho)||\tr_2 (\sigma)).
\end{equation} 
\end{theorem}
{\sf{Proof:}} In the case that $\ker \sigma\not\subseteq \ker \rho$ we have $ D(\rho||\sigma)=+\infty$
and the inequality (\ref{eq:mono-part-trace}) is trivially true.\\
Suppose that $\ker \sigma\subseteq \ker \rho $.
We will use the abbreviation $\tr$ for $\tr_{\cl(\hr_1)\otimes \cl(\hr_2)}$ and
$\tr_{\cl(\hr_1)}$ as well as for tensored versions thereof in what follows. It will always be clear from the context on which space the trace is acting.
Moreover we set
\[\rho_1:=\tr_2(\rho),\quad \sigma_1:=\tr_2(\sigma).  \]
Then for any $\eps\in (0,1)$ and all $n\in\nn$ we have
\begin{IEEEeqnarray}{rCl}\label{beta-rel}
 \beta_{\eps,n}(\rho_1,\sigma_2)&=& \min\{ \tr(\sigma_1^{\otimes n}a): a\in [0,\idn_{\hr_1^{\otimes n}}],
                                       \tr(\rho_1^{\otimes n}a)\ge 1-\eps   \}\nonumber\\
                              &=& \min\{\tr (\sigma^{\otimes n} E^{\otimes n}(a) ): a\in [0,\idn_{\hr_1^{\otimes n}}], 
                                          \tr(\rho^{\otimes n} E^{\otimes n}(  a ) )\ge 1-\eps   \}\nonumber\\
                              &=& \min\{\tr(\sigma^{\otimes n}a): a\in P_n: \tr(\rho^{\otimes n}a)\ge 1-\eps \}\nonumber\\ 
                              &\ge& \min\{\tr(\sigma^{\otimes n}a): a\in [0,\idn_{(\hr_1\otimes \hr_2)^{\otimes n}}], 
                                    \tr (\rho^{\otimes n}a)\ge 1-\eps\}\nonumber\\
                                      &=& \beta_{\eps,n}(\rho,\sigma),
\end{IEEEeqnarray}
where in the second line we have used the relation (\ref{partial-trace-2}), in the third line we used (\ref{partial-trace-4}),
while in the fourth line we used the elementary fact
 that the minimum value of a given function decreases if we enlarge the set we are minimizing over.\\
Taking $\log$ of both sides of (\ref{beta-rel}), dividing by $n$, and taking the limit shows, 
according to Theorem \ref{theorem-stein's-lemma},
that
\[ D(\rho_1||\sigma_2)\le D(\rho||\sigma) \]
as desired.\begin{flushright}$\Box$\end{flushright}
\begin{remark}
 The inequality (\ref{beta-rel}) has a nice intuitive interpretation: The minimum probability of error can only increase
if we have access to a smaller set of measurements upon which we can base our decisions.
\end{remark}
\begin{remark}
Notice that for the proof of Theorem \ref{mono-part-trace} we do not need the full power of Theorem \ref{theorem-stein's-lemma}. It is sufficient to
know that there is an increasing subsequence $(n_k)_{k\in \nn}$ of non-negative integers and a decreasing sequence $(\eps_{n_k})_{k\in \nn}$ with
$\eps_{n_k}\in (0,1)$ and $\lim_{k\to\infty}\eps_{n_k}=0$ such that
\[ \lim_{k\to\infty}\frac{1}{n_k}\log \beta_{\eps_{n_k},n_k}(\rho,\sigma)=-D(\rho,\sigma) \]
holds. 
\end{remark}
\begin{remark}\label{remark-general-mono}
 Theorem \ref{mono-part-trace} has a natural generalization to completely positive trace-preserving maps due to Lindblad \cite{lindblad-cptp-mono}: 
For any such map $T:\cl(\hr_1)\to\cl(\hr_2)$
and $\rho,\sigma\in \cs(\hr_1)$ we have
\begin{equation}\label{mono-cptp}
 D(\rho||\sigma)\ge D(T(\rho)||T(\sigma)).
\end{equation} 
The following generalization of the proof of Theorem \ref{mono-part-trace} to the situation of inequality (\ref{mono-cptp}) was suggested to us 
by Janis N\"otzel. 
We just have to replace the maps $\tr_2$ and $E$ by $T$ and its dual $T_{\ast}:\cl(\hr_2)\to\cl(\hr_1)$ defined via
\begin{equation}\label{mono-cptp-1}
 \tr(a T(a'))=\tr(T_{\ast}(a)a')
\end{equation} 
for $a\in \cl(\hr_2), a'\in\cl(\hr_1)$. The map $T_{\ast}$ is completely positive and unital, the latter meaning that $T_{\ast}(\idn_{\hr_2})=\idn_{\hr_1}$,
 implying that
for the set
\begin{equation}\label{mono-cptp-2}
 P_n:= T_{\ast}^{\otimes n}([0,\idn_{\hr_{2}^{\otimes n}}])
\end{equation} 
we have
\begin{equation}\label{mono-cptp-3}
 P_n\subseteq [0,\idn_{\hr_{1}^{\otimes n}}].
\end{equation} 
Then a similar reasoning as in the inequality chain (\ref{beta-rel}) leads to
\begin{equation}\label{mono-cptp-4}
 \beta_{\eps,n}(T(\rho),T(\sigma))\ge \beta_{\eps,n}(\rho,\sigma)
\end{equation} 
for all $n\in\nn$ and all $\eps\in (0,1)$ and we obtain (\ref{mono-cptp}) via Stein's lemma.
\end{remark}

Our next step is to show that the monotonicity of the relative entropy under partial trace, Theorem \ref{mono-part-trace},
implies the joint convexity of the relative entropy. To this end we extend slightly the definition of the relative entropy
to positive semi-definite operators and show that Theorem \ref{mono-part-trace} carries over to this generalized situation. 
For any pair of positive semi-definite operators $a,b\in \cl(\hr)$ we set 
\begin{equation}\label{def-rel-ent-2}
D(a||b):=
 \begin{cases}
  \tr(a\log a-a\log b) & \text{if }  \ker b\subseteq \ker a \\
  +\infty &   \text{else}.
 \end{cases}
\end{equation} 
\begin{corollary}[Monotonicity under partial trace II]\label{mono-part-trace-generalized}
 For any pair $a,
 b \in \cl(\hr_1)\otimes \cl(\hr_2)$ of positive semi-definite operators we have
\begin{equation}
D(a||b)\ge D(\tr_2(a)||\tr_2(b)).
\end{equation}  
\end{corollary}
{\sf{Proof:}} First note that in the case $\ker b \not\subseteq\ker a$ there is nothing to prove since $D(a||b)=+\infty$.\\
Let us suppose that $\ker b\subseteq \ker a$ holds. If $a=0$ there is again nothing to prove because
\begin{equation}\label{gen-mono-1}
 D(a||b)=0=D(\tr_2(a)||\tr_2(b)).
\end{equation} 
So, we can assume that $a\neq 0$ and consequently $b\neq 0$. Some simple algebra shows that for $\alpha,\beta\in \rr$, $\alpha,\beta>0$
it holds that
\begin{equation}\label{gen-mono-2}
 D(\alpha\cdot a||\beta \cdot b)=\alpha D(a||b)+ \tr(a)\alpha\log \frac{\alpha}{\beta}.
\end{equation} 
On the other hand, it follows from the definition of the partial trace that
\begin{equation}\label{gen-mono-3}
 \tr_{\cl(\hr_1)\otimes \cl(\hr_2)} (a)=\tr_{\cl(\hr_1)}(\tr_2(a))
\end{equation} 
for all positive semi-definite $a\in \cl(\hr_1)\otimes \cl(\hr_2)$.\\
Now we set $\alpha:=\frac{1}{\tr_{\cl(\hr_1)\otimes \cl(\hr_2)}(a)}, \beta:=\frac{1}{\tr_{\cl(\hr_1)\otimes \cl(\hr_2)}(b)}$ and obtain
from Theorem \ref{mono-part-trace}
\begin{equation}\label{gen-mono-4}
 D(\alpha \cdot a|| \beta\cdot b)\ge D(\alpha\cdot\tr_2(a)|| \beta\cdot  \tr_2(b)).
\end{equation} 
Taking into account (\ref{gen-mono-2}) and (\ref{gen-mono-3}) leads to
\begin{equation}\label{gen-mono-5}
 D(a||b)\ge D(\tr_2(a)||\tr_2(b))
\end{equation} 
and we are done. \begin{flushright}$\Box$\end{flushright}
\begin{corollary}[Joint convexity \cite{lindblad}]\label{theorem-joint-convexity}
 Let $\hr$ be a finite-dimensional Hilbert space over $\cc$, $a_1,\ldots,a_k,b_1,\ldots,b_k\in\cl(\hr)$ 
 positive semi-definite operators, and $\lambda_1,\ldots,\lambda_k\in \rr_{+}$ 
with $\sum_{i=1}^{k}\lambda_i=1$. Then
\begin{equation}\label{eq:joint-convexity}
 D\left( \sum_{i=1}^{k}\lambda_i a_i ||\sum_{i=1}^{k}\lambda_i b_i   \right)\le \sum_{i=1}^{k}\lambda_i D(a_i|| b_i).
\end{equation} 
\end{corollary}
{\sf{Proof:}} If for some $i\in \{1,\ldots ,k  \}$ $\ker b_i\not\subseteq \ker a_i$ then the right hand side of (\ref{eq:joint-convexity})
equals $+\infty$ and (\ref{eq:joint-convexity}) holds.\\
We may, therefore, suppose that for all $i\in\{1,\ldots, k  \}$ $\ker b_i\subseteq \ker a_i$ and define positive 
semi-definite operators $a',b'\in \cl(\hr)\otimes \cl(\cc^k)$ by
\[a':=\sum_{i=1}^{k}\lambda_i a_i\otimes |e_i\rangle\langle e_i|,\quad b':=\sum_{i=1}^{k}\lambda_i b_i\otimes |e_i\rangle\langle e_i|,  \]
where $\{e_1,\ldots, e_k  \}$ is an orthonormal basis of $\cc^k$. Note that $\ker b'\subseteq \ker a'$ and it is not hard to see that
\begin{equation}\label{joint-convexity-1}
 D(a'||b')= \sum_{i=1}^{k}\lambda_i D(a_i||b_i)
\end{equation} 
holds. Let $\tr_2: \cl(\hr)\otimes \cl(\cc^k)\to \cl(\hr)$ denote the partial trace. Then
\begin{equation}\label{joint-convexity-2}
\tr_2(a')=\sum_{i=1}^{k}\lambda_i a_i, \quad \tr_2(b')=\sum_{i=1}^{k}\lambda_i b_i,  
\end{equation}
and Corollary \ref{mono-part-trace-generalized} shows that
\begin{equation}\label{joint-convexity-3}
 D(a'||b')\ge D(\tr_2(a')||\tr_2(b')),
\end{equation} 
which is nothing else than (\ref{eq:joint-convexity}) by (\ref{joint-convexity-1}) and (\ref{joint-convexity-2}).
\begin{flushright}$\Box$\end{flushright}

\section{Lieb's concavity theorem: Tropp's argument}\label{sec:tropp}
In this section we shall outline Tropp's argument \cite{tropp} leading to the following theorem of Lieb \cite[Theorem~6]{lieb}:
\begin{theorem}\label{lieb-concavity}
 Let $\hr$ be a finite dimensional Hilbert-space over $\cc$ and let $h\in\cl(\hr)$ be self-adjoint. 
 Then the map $a\mapsto \tr \exp (h+\log a)$ is concave on the positive-definite cone of
$\cl(\hr)$.
\end{theorem}
Tropp's proof of Theorem \ref{lieb-concavity} is based on a sequence of lemmas which we will present first.
\begin{lemma}\label{klein-variational}
 1. (Klein's Inequality \cite{klein}) Let $a,b\in \cl(\hr)$ be positive semi-definite operators. Then
 \begin{equation}\label{klein-inequality}
  D(a||b)\ge \tr(a-b).
 \end{equation} 
2. (Variational Formula for Trace) For any positive-definite $b\in \cl(\hr)$ we have
\begin{equation}\label{variational-formula}
 \tr(b)= \max_{x\in\cl(\hr), x\ge 0}\tr(x\log b- x\log x + x).
\end{equation} 
\end{lemma}
{\sf{Proof:}} 1. We may suppose that $\ker b\subseteq \ker a$ since otherwise the inequality (\ref{klein-inequality})
is clearly true. Moreover we can assume that $a\neq 0$ because (\ref{klein-inequality}) is trivially satisfied in the case $a=0$. 
Since $a\neq 0$ implies $b\neq 0$ we see that $\tr(a),\tr(b)>0$ and Corollary \ref{mono-part-trace-generalized} applied with $\hr_1:=\cc$, $\hr_2=\hr$
shows that
\begin{equation}\label{eq:klein-1}
 D(a||b)\ge D(\tr(a)||\tr(b))=\tr (a)\log \frac{\tr(a)}{\tr(b)}\ge \tr(a)-\tr(b),
\end{equation} 
where the last inequality follows from the numerical inequality $\log x\ge -\frac{1}{x}+1$ valid for all positive numbers $x$.\\
2. Note that the inequality $\tr(b)\ge \tr(x\log b-x\log x+x)$ is nothing else than (\ref{klein-inequality}) and 
equality holds for $x=b$.\begin{flushright}$\Box$\end{flushright}
The final lemma we need for the proof of Theorem \ref{lieb-concavity} is Lemma 2.3 from \cite{carlen-lieb}. We omit the elementary proof.
\begin{lemma}\label{concave} 
Let $f: K_1\times K_2\to\rr$ be a jointly concave function such that for each $y\in K_2$ there is $x'\in K_1$ such that
\[ f(x',y)=\sup_{x\in K_1}f(x,y), \]
i.e. $\sup$ is attained for each $y$ and is in fact $\max$. Then the function $y\mapsto \max_{x\in K_1}f(x,y)$ is concave.
\end{lemma}
{\sf{Proof of Theorem \ref{lieb-concavity}}:} We apply Lemma \ref{klein-variational}.2 with $b:=\exp(h+\log a)$ and end up with
\begin{IEEEeqnarray}{rCl}\label{lieb-conc-1}
 \tr(\exp (h+\log a))&=& \max_{x\ge 0}\tr(x(h+\log a)-x\log x +x )\nonumber\\
                      &=& \max_{x\ge 0}(\tr(xh)-D(x||a)+\tr(x)).
\end{IEEEeqnarray} 
The right hand side of (\ref{lieb-conc-1}) is concave in $a$ by Lemma \ref{concave} due to the fact that the map
$(x,a)\mapsto \tr(xh)-D(x||a)+\tr(x)$ jointly concave for fixed $h$ by Corollary \ref{mono-part-trace-generalized}.
\begin{flushright}$\Box$\end{flushright}
\begin{remark}
 It is an interesting aside to have a look at Lindblad's proof \cite{lindblad} of the joint convexity of the relative entropy 
(which Tropp \cite{tropp} cites) and Tropp's argument as a whole.\\
Lindblad's starting point is a special case of \cite[Theorem~1]{lieb} stating that on pairs of positive-semidefinite operators the map 
$(a,b)\mapsto\tr(a^{1-p}b^p )$ 
is jointly concave for any $p\in[0,1]$. He then observes that the derivative of that map with respect to $p$ at $p=0$ is $-D(a||b)$ from which
the joint convexity of the relative entropy follows. Tropp shows how to derive the concavity of $a\mapsto \tr\exp(h+\log a)$ ($h$ self-adjoint)
on the positive cone from the joint convexity of the relative entropy. Thus, when seen in a sequence, the arguments of Lindblad and Tropp
show in few lines that the joint concavity of $(a,b)\mapsto\tr(a^{1-p}b^p )$  ($p\in [0,1]$, $a,b$ positive-semidefinite) 
implies the concavity of $a\mapsto \tr\exp(h+\log a)$ ($h$ self-adjoint, $a$ positive-definite).\\
In a similar vein, following the proofs of Corollaries 2.1 or 2.1 in Effros' paper \cite{effros} we can see that Theorem \ref{lieb-concavity} can be 
easily deduced from the operator convexity of $f(x)=x\log x$ or $g(x)=-x^p$, for $p\in [0,1]$. 
\end{remark}

\section{Historical remarks and related work}\label{sec:history}
Stein's lemma in the classical form appears for the first time in Chernoff's (!) work \cite{chernoff}. In the quantum realm, 
 Hiai and Petz \cite{hiai-petz}
have shown that for all $\eps\in (0,1)$
\[ \limsup_{n\to\infty}\frac{1}{n}\log \beta_{\eps,n}(\rho,\sigma)\le -D(\rho||\sigma),\]
and 
\begin{equation}\label{hiai-petz-converse}
\liminf_{n\to\infty}\frac{1}{n}\log \beta_{\eps,n}(\rho,\sigma)\ge -\frac{D(\rho||\sigma)}{1-\eps}
\end{equation}
hold. The proof of the inequality (\ref{hiai-petz-converse}) presented in \cite{hiai-petz} relies on the monotonicity of the quantum relative entropy.
Ogawa and Nagaoka \cite{ogawa} obtained the strong converse, i.e. they showed that $1-\eps$ on the right hand side of (\ref{hiai-petz-converse})
can be replaced by $1$ thus leading to
\[ \lim_{n\to\infty}\frac{1}{n}\log \beta_{\eps,n}(\rho,\sigma)= -D(\rho||\sigma) \]
for all $\eps\in (0,1)$. The proof in \cite{ogawa} relies on the monotonicity of quantum quasi-entropies which present
a generalization of relative entropy \cite{petz-1}, \cite{petz-2}.\\
The monotonicity of the relative entropy under the partial trace (MPT) has been shown by Lieb and Ruskai in \cite{lieb-ruskai}.
Their proof  relies on the concavity of $a\mapsto \tr (\exp(h+\log a ))$ ($h$ self-adjoint) on the positive cone of $\cl(\hr)$ which 
was proven by Lieb in \cite{lieb}. Uhlmann \cite{uhlmann}
observed that this implies so called strong subadditivity of von Neumann entropy. The paper \cite{lieb-ruskai} by Lieb and Ruskai, in turn, 
contains an argument showing
that the strong subadditivity of von Neumann entropy implies MPT.\\
Joint convexity of the relative entropy was established by Lindblad \cite{lindblad}. Lindblad's proof uses another theorem of Lieb \cite{lieb}
which states that the map $(a,b)\mapsto \tr(a^{1-p}b^p)$, $p\in [0,1]$, is jointly concave in $(a,b)$ for positive semi-definite $a,b\in \cl(\hr)$.
The monotonicity of the quantum relative entropy under the action of completely positive trace-preserving maps was established by Lindblad in 
\cite{lindblad-cptp-mono}.
Uhlmann \cite{uhlmann-wyd-concavity} derives the monotonicity as well as the joint convexity of the relative entropy in the general setting of operator
algebras via interpolation theory.
An ingenious analytic proof of joint convexity of the relative entropy is discovered by Simon \cite[Ch.~8]{simon}.\\
In \cite{effros} Effros gives very short and elegant proofs of joint convexity of relative entropy and several results of Lieb from \cite{lieb} based on
the notion of operator convex functions and Jensen's inequality for operators proven by Hansen and Pedersen \cite{hansen-pedersen}.\\
More historical facts of interest as well as other analytic approaches to the properties of quantum relative entropy and interrelation among the entropy
inequalities can be picked up in the 
nice review \cite{ruskai} by Ruskai.\\
Finally, note that the ansatz to derive inequalities for matrices or entropy from operational, information theoretic, or probabilistic
 interpretation of the quantities 
in question is not new at all. Already Dembo, Cover, and Thomas in \cite{dembo-cover-thomas} derived several matrix inequalities from the properties 
of multivariate
gaussian distributions.\\
Much closer in spirit to our work is Winter's \cite{winter} derivation of the famous Holevo bound \cite{holevo} from the coding theorem with the 
strong converse for channels with classical input and quantum mechanical output.

\vspace{0.3cm}
\noindent
{\sf\textbf{Acknowledgement}} We thank Holger Boche, Gisbert Janßen, Janis Nötzel, and Moritz Wiese for encouragement to publish this work and 
many discussions, comments, and suggestions during the process of the preparation of the manuscript. We thank especially Janis N\"otzel for sharing
with us his proof of the monotonicity under completely positive trace-preserving maps reproduced in Remark \ref{remark-general-mono}.
\\
While preparing the first version \cite{mono-1} of the present paper we had several discussions with 
Tyll Kr\"uger, Ruedi Seiler, Arleta Szko\l a, and Andreas Winter and we thank them for that. Mary Beth Ruskai contributed several
remarks and corrections concerning the history of and interrelation between the entropy inequalities.\\
The support by the Deutsche Forschungsgemeinschaft (DFG) via grant BO 1734/20-1 and Bundesministerium für Bildung und Forschung (BMBF) via grant
01BQ1050 is gratefully acknowledged.

\appendix
\section{Proof of Lemma \ref{overlap-lemma}}\label{proof-overlap-lemma}
The proof relies on the following simple fact: For all $a,b\in\cl(\kr)$ we have
\begin{equation}\label{a-overlap-1}
 |\tr(a^{\ast}b)|=|\tr(b^{\ast}a)|\le \sqrt{\tr(a^{\ast}a)}\sqrt{\tr(b^{\ast}b)},
\end{equation} 
where the equality stems from the fact that $\tr(x^{\ast})=\overline{\tr(x)}$ and the inequality is nothing
else than the Cauchy-Schwarz inequality for the Hilbert Schmidt inner product $(a,b)\mapsto \langle a,b\rangle_{HS}:=\tr(a^{\ast}b)$ on the space
$\cl(\kr)$.\\
In what follows we will write $\idn$ for $\idn_{\kr}$. For $p,q$ as given in the statement of the lemma we have
\begin{equation}\label{a-overlap-2}
 0\le (\idn-p)q(\idn-p)=-q +q(\idn-p)+(\idn-p)q+pqp,
\end{equation}
which is readily verified. Multiplying (\ref{a-overlap-2}) from left and from right by $\tau^{1/2}$, taking trace of both sides, and rearranging
leads to
\begin{IEEEeqnarray}{rCl}\label{a-overlap-3}
 \tr(\tau q)&\le & \tr(\tau^{1/2} q(\idn-p)\tau^{1/2})+\tr(\tau^{1/2}(\idn-p)q\tau^{1/2})+\tr(\tau pqp)\nonumber\\
            &=& |\tr(\tau^{1/2}q(\idn-p)\tau^{1/2})|+|\tr(\tau^{1/2}(\idn-p)q\tau^{1/2})|+\tr(\tau pqp)\nonumber\\
            &=& 2 |\tr(\tau^{1/2}(\idn-p)q\tau^{1/2})|+\tr(\tau pqp)\nonumber\\
            &\le& 2 \sqrt{\tr(\tau(\idn-p)^{2})}\sqrt{\tr(\tau q^2)} +\tr(\tau pqp)\nonumber\\
            &\le & 2\sqrt{\tr(\tau(\idn-p))} +\tr(\tau pqp),
\end{IEEEeqnarray}  
where in the third line we have used the left relation in (\ref{a-overlap-1}), in the fourth we used the Cauchy-Schwarz inequality 
(right half of (\ref{a-overlap-1})) with $a^{\ast}=\tau^{1/2}(\idn-p), b=q\tau^{1/2}$, while in the last line we estimated
$(\idn-p)^2\le \idn-p$ since $0\le \idn-p\le \idn$, and $\tr(\tau q^2)\le \tr(\tau q)\le 1$.\\
This shows
\begin{equation}\label{a-overlap-4}
 \tr (\tau pqp)\ge \tr(\tau q)-2\sqrt{\tr (\tau (\idn-p))}.
\end{equation} 
For the second part of the lemma we simply observe that
\begin{IEEEeqnarray}{rCl}\label{a-overlap-5}
 \tr(pqp)&=& \tr (upqp )+\tr((\idn-u)pqp)\nonumber\\
         &\ge& \tr(upqp)\qquad (\textrm{since  } \tr((\idn-u)pqp)\ge 0)\nonumber\\
         &\ge&\frac{1}{c}\tr (\tau u pqp)\qquad (\textrm{since } \tau u\le c u)\nonumber\\
         &=& \frac{1}{c}( \tr (\tau pqp)- \tr(\tau (\idn-u)(pqp)) )\nonumber\\
         &\ge& \frac{1}{c}(\tr(\tau pqp)-\tr (\tau (\idn-u)) ),
\end{IEEEeqnarray}
in the last line we have used the fact that $pqp\le \idn$. Now, a combination of (\ref{a-overlap-5})
and (\ref{a-overlap-4}) finishes the proof of the second part of the lemma.
\section{Singular case of Stein's lemma}
In this appendix we provide the variant of Theorem \ref{theorem-stein's-lemma} for the case $\ker\sigma\not\subseteq \ker \rho$.
\begin{lemma}\label{singular-stein}
 Let $\rho,\sigma\in \cs(\hr)$ with $\ker\sigma\not\subseteq \ker \rho$. Then to each $\eps\in (0,1)$ there is $n_0(\eps)\in \nn$ such that
for all $n\in\nn, n\ge n_0(\eps)$
\begin{equation}\label{eq:singular-stein}
 \beta_{\eps,n}(\rho,\sigma)=0.
\end{equation} 
\end{lemma}
{\sf{Proof:}} Since $\ker\sigma\not\subseteq \ker \rho $ there is $e\in \ker \sigma$, $e\notin \ker \rho$ with $||e||=1$. Then, clearly,
\begin{equation}\label{eq:singular-stein-1}
 \tr (\sigma |e\rangle\langle e |)=0, \qquad \tr(\rho |e\rangle\langle e|)=a\in (0,1].
\end{equation} 
We set
\begin{equation}\label{eq:singular-stein-2}
 p_0:=|e\rangle\langle e|,\qquad p_1:=\idn_{\hr}-p_0.
\end{equation} 
For $n\in \nn$ we introduce
\begin{equation}\label{eq:singular-stein-3}
 T_n:=\{ x^n\in \{ 0,1 \}^n: x^n \textrm{ contains at least one }0 \},
\end{equation} 
and the projection
\begin{equation}\label{eq:singular-stein-4}
 q_n:=\sum_{x^n\in T_n}p_{x^n}
\end{equation} 
where $p_{x^n}:=\otimes_{i=1}^{n}p_{x_{i}}$. 
Notice that
\begin{equation}\label{eq:singular-stein-5}
 q_n+p_1^{\otimes n}=\idn_{\hr^{\otimes n}}=\idn_{\hr}^{\otimes n}
\end{equation} 
which leads to
\begin{equation}\label{eq:singular-stein-6}
 \tr(\rho^{\otimes n}q_n)=\tr(\rho^{\otimes n}(\idn_{\hr}^{\otimes n}-p_1^{\otimes n}))=1-(1-a)^n
\end{equation} 
by (\ref{eq:singular-stein-1}).
On the other hand it is clear by the definition of $q_n$  and $T_n$ and by (\ref{eq:singular-stein-1}) that
\begin{equation}\label{eq:singular-stein-7}
 \tr(\sigma^{\otimes n}q_n)=0.
\end{equation}  
Consequently, by (\ref{eq:singular-stein-6}) and (\ref{eq:singular-stein-7}) for any $\eps\in (0,1)$ there is $n_0(\eps)\in\nn$ such
that for all $n\in\nn, n\ge n_0(\eps)$  
\begin{equation}\label{eq:singular-stein-8}
 \beta_{\eps,n}(\rho,\sigma)\le \tr(\sigma^{\otimes n}q_n)=0
\end{equation} 
\begin{flushright}$\Box$\end{flushright}

\end{document}